%% file: DLPKDD_GPatch.tex
\documentclass[sigconf]{acmart}
\input{math_commands.tex}

\theoremstyle{definition}

\usepackage{enumitem}

\AtBeginDocument{%
	\providecommand\BibTeX{{%
			\normalfont B\kern-0.5em{\scshape i\kern-0.25em b}\kern-0.8em\TeX}}}

\acmConference[DLP-KDD 2022]{4th Workshop on Deep Learning Practice and Theory for High-Dimensional Sparse and Imbalanced Data with KDD 2022}{August 14, 2022}{Virtual Conference, Online}
\acmPrice{15.00}
\acmISBN{978-1-4503-XXXX-X/22/08}

\begin{document}
	\author{Hao Chen}
	\email{sundaychenhao@gmail.com}
	\affiliation{%
		\institution{Tencent Inc.}
		\state{}
		\country{}
	}

	\author{Zefan Wang}
	\email{wongzfn@gmail.com}
	\affiliation{%
		\institution{Jinan University.}
		\state{}
		\country{}
	}

	\author{Yue Xu}
	\email{yuexu.xy@foxmail.com}
	\affiliation{%
		\institution{Alibaba Group.}
		\state{}
		\country{}
	}
	
	\author{Xiao Huang}
	\email{xiaohuang@comp.polyu.edu.hk}
	\affiliation{%
		\institution{The Hong Kong Polytechnic University.}
		\state{}
		\country{}
	}
	
	\author{Feiran Huang}
	\email{huangfr@jnu.edu.cn}
	\affiliation{%
		\institution{Jinan University.}
		\state{}
		\country{}}
	
	\title{GPatch: Patching Graph Neural Networks for Cold-Start Recommendations}
	


	
	\begin{abstract}
	Cold start is an essential and persistent problem in recommender systems. State-of-the-art solutions rely on training hybrid models for both cold-start and existing users/items, based on the auxiliary information. Such a hybrid model would compromise the performance of existing users/items, which might make these solutions not applicable in real-worlds recommender systems where the experience of existing users/items must be guaranteed. Meanwhile, graph neural networks (GNNs) have been demonstrated to perform effectively warm (non-cold-start) recommendations. However, they have never been applied to handle the cold-start problem in a user-item bipartite graph. This is a challenging but rewarding task since cold-start users/items do not have links. Besides, it is nontrivial to design an appropriate GNN to conduct cold-start recommendations while maintaining the performance for existing users/items. To bridge the gap, we propose a tailored GNN-based framework~(GPatch) that contains two separate but correlated components. First, an efficient GNN architecture --- GWarmer, is designed to model the warm users/items. Second, we construct correlated Patching Networks to simulate and patch GWarmer by conducting cold-start recommendations. Experiments on benchmark and large-scale commercial datasets demonstrate that GPatch is significantly superior in providing recommendations for both existing and cold-start users/items.
	
	\end{abstract}
	
	
	
	\maketitle
	
	\section{Introduction}
    Recommender systems aim to retrieve deserving information for users out of massive online content and provide personalized recommendations. There are thousands of user registrations and item productions in various recommender systems every second, introducing cold-start problems. That is how to recommend for the users and items without historical interactions. In a typical user-item cold-start recommendation scene, recommender systems can access the auxiliary information for both existing and cold-start users/items to alleviate the cold-start problem.
    
    Considering the large proportion of existing users and items, one main challenge of cold-start recommendation is to serve cold-start recommendations without affecting the recommendations for existing users and items. Given trained warm embeddings and the auxiliary information, state-of-the-art solutions such as DropoutNet~\cite{volkovs2017dropoutnet} and Heater~\cite{zhu2020heater} train hybrid models to recommend for both cold-start users/items and existing users/items. Such hybrid models have to tackle two totally different types of inputs simultaneously: 1. for warm~(non-cold-start) recommendations, feeding with \textit{warm embeddings} and the auxiliary information to generate the warm representation; and 2. for cold-start recommendations, feeding with \textit{vacant embeddings} and the auxiliary information to generate the cold-start representation. Although showing certain cold-start ability, such hybrid cold-start models would compromise the performance of existing users/items due to accommodating the two entirely different inputs: vacant embeddings and the trained warm embeddings. The hybrid models cannot provide satisfying recommendations for existing users and items, restricting these solutions cannot be applied in real-world recommender systems, where the experience of existing users/items must be guaranteed.

    The core of the warm recommender system is how to model the interactions of existing users and items. Accordingly, Graph Convolutional Networks~(GCNs), skilled in modeling graph-structured data, have achieved the state-of-the-art performance in many recommendation applications~\cite{fu2020magnn,chen2022neighbor,lu2020meta,chen2021non,liu2020heterogeneous,chen2020label,fan2019meirec,zhao2019intentgc,chen2022generative,chen2020graph}. NGCF~\cite{wang2019ngcf} and LightGCN~\cite{he2020lightgcn} construct bipartite graphs from historical user-item interactions and then embed the users and items with the help of graph convolutional networks~\cite{kipf2016gcn}. However, current GCN models cannot be directly applied to deal with the cold-start problem  due to three major challenges as follows. First, there are no historical interactions for the cold-start users/items. It means that they do not have neighbors. GCN models could not perform neighborhood aggregations to refine the embedding representations of cold-start users/items. They do not have exact representations either. Second, it is nontrivial to design a suitable GCN to maintain the performance of recommendation on existing users/items, while performing cold-start recommendations. Current cold-start models are designed to tackle the non-graph recommendation models. Third, due to the recursive message passing structure of GCN models, GCN models are much time consuming than simple inner-product functions. Performing cold-start GCN models may encounter computational burdens than the current inner-product-based cold-start models.


	To overcome the above challenges, this paper proposes Graph Patching Networks~(GPatch) to provide satisfying recommendations for both existing and cold-start users and items. GPatch contains two separative but correlated models: GWarmer and Patching Networks. GWarmer is a specially designed efficient GCN Model which enhances the recommendation for existing users/items. It is impressive that GWarmer is as simple as the inner product during inferring, making GPatch more efficient than SOTA cold-start models. Besides, serving as patches for GWarmer, Patching Networks afford cold-start recommendations by generating GWarmer-friendly embeddings from the auxiliary information. Unlike traditional cold-start models, Patching Networks only serve cold-start recommendations and thus will not influence the warm recommendation models. Extensive experiments on two benchmark and one 15-day WeChat commercial datasets, with three different type of embedding methods, present that GPatch statistically significantly outperforms other recommendation models on both hybrid recommendation performance~(including warm/cold-start recommendation tasks) and pure-warm recommendation performance. 
		\begin{figure}[tbp]
		\centering
		\includegraphics[width=1.01\columnwidth , trim=1.6cm 0.4cm 1.2cm 1.7cm,clip]{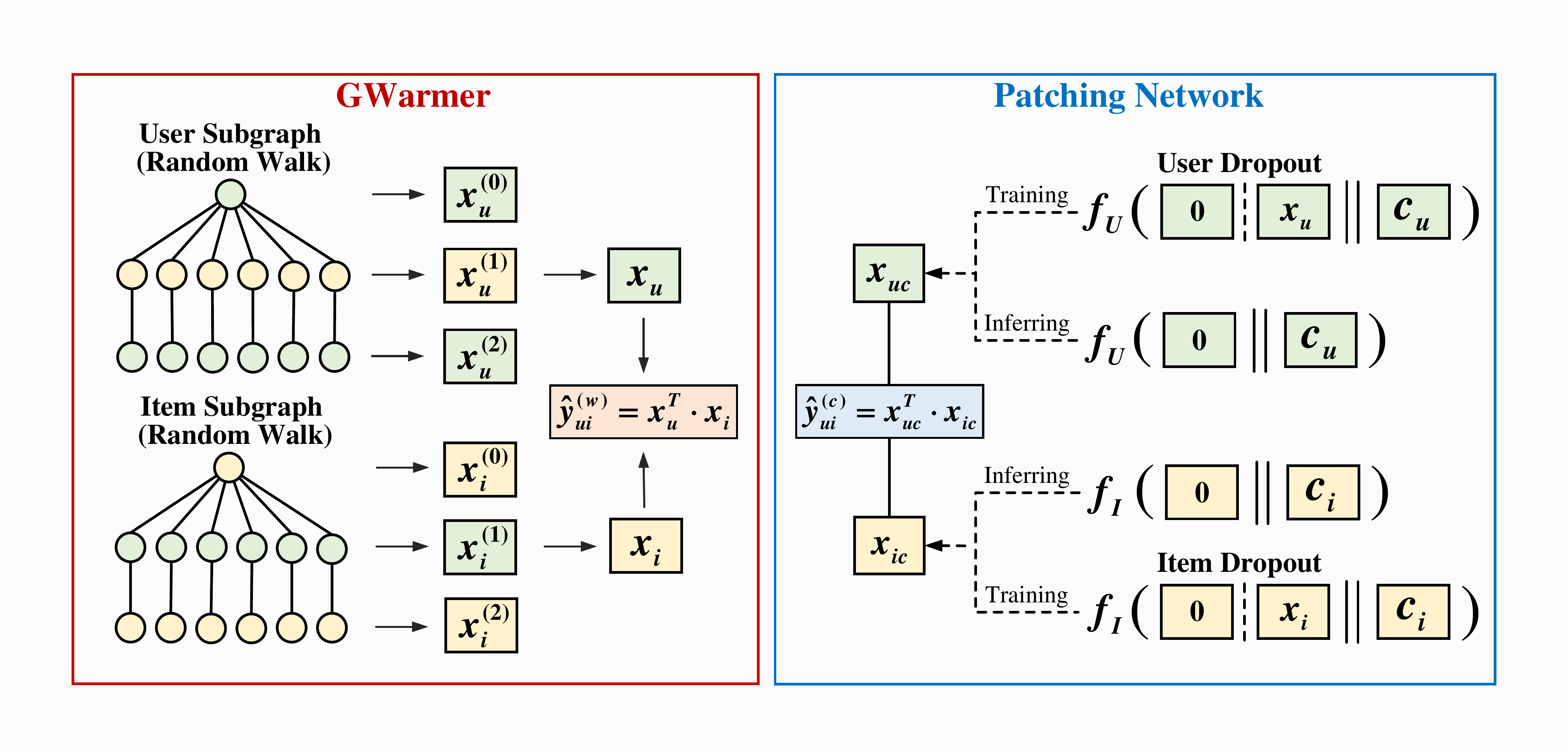}
		\vspace{-3em}
		\caption{Sketch of GPatch. GWarmer provides recommendation for existing users and items, while Patching Networks recommend for cold-start users and items.}
	\vspace{-2em}
	\label{fig:framework}
	\end{figure}
    In summary, our contributions are listed as below:
	\begin{itemize}
		\item This paper at the first time enables the cold-start recommendation of GCN models, resulting in better recommendation performance for both existing/cold-start users and items.
		\item This paper first notices the compromising caused by the hybrid cold-start models and proposed Patching Networks that serve cold recommendations without compromising the performance of existing users/items.
		\item Experiments on two benchmark and one online commercial datasets, with three different embeddings, show that GPatch significantly outperforms other cold-start models on all situations with statistically significant margins.
	\end{itemize}

	\section{Our Method: GPatch}
	This section first presents the whole framework of our GPatch, and then we introduce the details of the proposed GWarmer and Patching Networks.
	
	\subsection{Overall Framework}
	Let $u\in\gU$ denote the user index, and $i\in\gI$ denote the item index. $\gG$ denotes the user-item bipartite graph constructed from the user-item interactions $\gR$. $\gU_w$ and $\gI_w$ denote the existing~(warm) user/item set. $\mC_U$ and $\mC_I$ denote the auxiliary information for all users/items. $\mE_U$ and $\mE_I$ denote the user/item embedding matrix trained by any given user-item embedding model.
    
    We first define the general form of the warm/cold-start recommendation task. \emph{Warm recommendations}, by definition, mean that recommend warm items to warm users. For warm users and items, we can access the warm embeddings $\mE_U$ and $\mE_I$ and the historical interaction graph $\gG$. Formally, GWarmer predicts the relevance of a warm user-item pair $(u,i)$ with $\gW(u,i;\gG,\mE_U,\mE_I)$. By contrast, \emph{Cold-start Recommendations} refer to the recommendation for either cold-start user or cold-start item, since the recommendation model cannot access the historical interaction graph and the embeddings for both users and items simultaneously. Without the graph and the embeddings, Patching Networks generate their cold-start embeddings from their auxiliary information $\vc_u$ and $\vc_i$, namely, $\gP(u,i;\vc_u,\vc_i)$. After defining GWarmer and Patching Networks, for an arbitrary user-item pair, GPatch computes its relevance scores with the following formula,
	\begin{align}
	\hat{y}_{ui} = 
		\left\{
		\begin{aligned}
		&\gW(u,i;\gG,\mE_U,\mE_I) , \quad &u\in \gU_w \text{ and } i \in \gI_w , \\
		&\gP(u,i;\vc_u,\vc_i), &u \notin \gU_w \text{ or } i \notin \gI_w,
		\end{aligned}
		\right.
	\end{align}
	where $\gW(\cdot)$ denotes GWarmer, and $\gP(\cdot)$ denotes Patching Networks. We will detail these two parts in the next two subsections.
	\subsection{GWarmer}
    \label{sec:gwarmer}
	Common models~\cite{rendle2012bprmf,rao2015grmf,huang2017label,he2020lightgcn} usually focus on investigating more powerful user/item embeddings and then predict the relevance scores with inner product function. 
	\begin{equation}
	    \hat{y}_{ui}^{(w)} = \ve_u^\top \ve_i.
	    \label{eq:inner}
	\end{equation}
	
	Correspondingly, an ideal warm recommendation model for real-world recommender systems, where user-item embeddings have already been trained by various warm embedding models, should be able to enhance any given embeddings with graph structure $\gG$, without specifying the embedding models~(such as MF~\cite{koren2009MF}, MetaPath2Vec(M2V)~\cite{dong2017metapath2vec} or LightGCN~\cite{he2020lightgcn}). Besides, the warm recommendation model should be efficient in training and inferring.
	
	As illustrated in~\autoref{fig:framework}, GWarmer first constructs the $K$-layer subgraphs for warm users/items with random-walk and then aggregates the random-walk set to represent the graph topology of warm users/items. Finally, a self-adaptive inner product relevance function is designed to perform the warm recommendation.
	
	\textbf{Random-Walk Step.} Random-walk is an efficient method in extracting the graph topology~\cite{perozzi2014deepwalk,grover2016node2vec}. As shown in~\autoref{fig:framework}, for any given warm user/item $t$, we repetitively generate $K$-length walks $S$ times with setting $t$ as the root node, where $S$ denotes the sampling size. Here we denote the random-walk set as $\gT_K(t)$. Note that in bipartite user-items graphs, user nodes connect to item nodes while item nodes connect to user nodes. In~\autoref{fig:framework}, we use green circles to denote user nodes and yellow circles to denote item nodes.
	
	\textbf{Walk Pooling Step.} After obtaining the random-walk set, we substitute the node ids by their corresponding embedding vectors, then we aggregate the random-walk set with mean pooling aggregator~\cite{gunets}. Mean pooling is permutation-invariant on set elements, such that the order of the walks will not influence the aggregation. After the mean pooling aggregation, for node $t$ we get its layer-wise representation for each layer of the subgraph. Formally, the layer-wise representation can be calculated with the following formula,
	\begin{equation}
	    \vx_t^{(k)} =\frac{1}{S}\sum_{\mT_i\in\mathcal{T}_K(v)}\mT_{i}^{(k)},
	\end{equation}
	where $k$ denotes the layer number of the subgraph and $S$ denotes the sampling size of the random-walk. Specifically, $\vx_t^{(0)}=E_t$.
	
    \textbf{Self-adaptive Inner Product} With the above two steps, we have $K$ representation vectors for each warm user or item $t$. Taken user $u$ as an example, $\vx_u^{(0)}$  is her own embeddings, while $\vx_u^{(1)}$ contains the mean embedding of her interacted items, and $\vx_u^{(2)}$ aggregates the embeddings of the users that have shared interacted items with $u$. From this point of view, the representations of different layers profile user $u$ from various aspects. Motivated by this, we introduce the self-adaptive weighted sum which assigns different weights for different layers. This further enhances the flexibility and extensibility of the inner product. By specifying the user representations and item representations with different weights, the relevance function is given as,
    \begin{gather}
    	\vx_u = \sum_k w_u^{(k)}\vx_u^{(k)}, \  \vx_i = \sum_k w_i^{(k)}\vx_i^{(k)}, \label{eq:warm}\\
    	\hat{y}_{ui}^{(w)} =  \gW(u,i;\gG,\mE_U,\mE_I) = \vx_u^\top\vx_i,
    \end{gather}
    where $\{w_u^{(0)},\cdots,w_u^{(K)}\}$ and $\{w_i^{(0)},\cdots,w_i^{(K)}\}$ denotes the self-adaptive weights for users' and items' layer-wise representations, respectively. $\vx_u$ and $\vx_i$ denotes the GWarmer representation of $u$ and $i$.
    
	Comparing with other recursive message passing models, GWarmer is much suitable for large-scale online recommender systems. First, the random walk step and the walk pooling step are parameter-free, which could be pre-computed and pre-stored offline in a parallel manner. Second, in large commercial systems, the historical user-item interactions are stored in distributed databases. Comparing with traditional recursive GCNs which cost an extensive query time for the recursive neighbor expanding, random walk policy needs significantly lower I/O times. Finally, for the online inferring process, after storing the GWarmer representations, GWarmer is as simple as inner product functions, which can be easily adapted with online recommender frameworks~(\emph{e.g.}, Faiss~\cite{johnson2019faiss}).

	\subsection{Patching Networks}
	Severing as patches of GWarmer, Patching Networks have to be consistent with GWarmer when recommending for cold-start recommendation tasks. To this end, as shown in~\autoref{fig:framework}, we bridge GWarmer and Patching Networks by using GWarmer representation instead of raw embeddings to train the Patching Networks.
    As shown in~\autoref{fig:framework}, we employ the popular cold-start dropout mechanism~\cite{volkovs2017dropoutnet,zhu2020heater} to optimize the Patching Networks by randomly masking the GWarmer representations of warm users/items with ratio $\tau$. Given any warm node $t$, we sample a random variable $p\in\{0,1\}$ from $Bernoulli(\tau)$. Then the masked GWarmer representation is:
	\begin{equation}
	    \gD(\vx_t, \tau) =  p\cdot \mathbf{0} +  (1-p)\cdot \vx_t,
	\end{equation}
	where $\mathbf{0}$ is a zero vector that has the same dimension as $\vx_t$.
	
	Given the warm user-item pair $(u,i)$, the Patching Networks predict its relevance scores with following formula,
	    \begin{gather}
    	\vx_{uc} = f_U(\gD(\vx_u,\tau)\parallel \vc_u), \ \vx_{ic} = f_I(\gD(\vx_i,\tau)\parallel \vc_i), \label{eq:patcher}\\
    	\hat{y}_{ui}^{(c)}=\gP(u,i;\vc_u,\vc_i) = \vx_{uc}^\top\vx_{ic},
    \end{gather}
    where $f_U$ and $f_I$ denotes the mapping function for users and items, respectively. Note that, $p$ is set as 1 during the inference of cold-start recommendation tasks when we set the embeddings of cold-start users/items as vacant embedding \textbf{0}.

    In GPatch framework, the self-adaptive weights in Eq.~(\ref{eq:warm}) and the mapping functions in Eq.~(\ref{eq:patcher}) require optimization. Thus, we employ the Mean Squared Error~(MSE) loss~\cite{volkovs2017dropoutnet,zhu2020heater} to optimize both GWarmer and Patching Networks  simultaneously. Formally, the loss function contains two parts and we present it as:
	\begin{equation}
	\gL = \sum_{\gR^+\cup\gR^-}(y_{ui} - \gW(u,i;\gG,\mE_U,\mE_I))^2 + (y_{ui} - \gP(u,i;\vc_u,\vc_i))^2,
	\end{equation}
	where $\gR^+$ denotes the observed user-item interactions, while $\gR^-$ denotes the sampling negative interactions. Thus, $y_{ui}=1$ for $(u,i) \in \gR^+$ and $y_{ui}=0$ for $(u,i) \in \gR^-$.
	\section{Experiments}

\begin{table}[tbp]
\small
\centering
\caption{Statistics of the three datasets}
\vspace{-1em}
\setlength{\tabcolsep}{2.7mm}{
 \begin{tabular}{ccccc}
 \toprule
 Dataset & User \#  & Item \# & Interaction \# & Density \\
 \midrule
 CiteULike & 5,551  & 16,980 & 204,986 & 0.22\% \\
 XING  & 106,881 & 20,519 & 3,856,580 & 0.18\% \\
 WeChat & 87,772 & 446,408 & 17,768,997 & 0.05\% \\
 \bottomrule
 \end{tabular}%
\vspace{-1em}
	}
	\label{tab:dataset}
\end{table}

\begin{table*}[ht]
\small
\centering
\vskip -1em
\caption{Comparison of hybrid recommendation performance. The percentages give the relative improvement over the best baseline (underlined). The superscript * denotes statistically significance against the best baseline with p<0.05.}
\vspace{-1em}
\resizebox{1\textwidth}{!}{
\setlength{\tabcolsep}{2mm}{
			
 \begin{tabular}{ccccccccccc}
 \toprule
 Embeddings & Models & CiteULike(REC@20) & CiteULike(PRE@20) & CiteULike(NDCG@20)  & XING(REC@20) & XING(PRE@20) & XING(NDCG@20)  & WeChat(REC@100) & WeChat(PRE@100) & WeChat(NDCG@100) \\
 \midrule
 MF+   & DropoutNet & 0.0780 & 0.0359 & 0.0654 & 0.1988 & 0.0712 & 0.1686 & 0.0519 & 0.0108 & 0.0315 \\
       & Heater & \underline{0.0968} & \underline{0.0479} & \underline{0.0849} & \underline{0.2190} & \underline{0.0792} & \underline{0.1921} & \underline{0.0640} & \underline{0.0130} & \underline{0.0403} \\
       & GPatch & \textbf{0.1404(45.04\%)}$^*$ & \textbf{0.0668(39.46\%)}$^*$ & \textbf{0.1187(39.81\%)}$^*$ & \textbf{0.2346(7.12\%)}$^*$ & \textbf{0.0866(9.34\%)}$^*$ & \textbf{0.2055(6.98\%)}$^*$ & \textbf{0.0659(2.97\%)}$^*$ & \textbf{0.0133(2.31\%)}$^*$ & \textbf{0.0443(9.93\%)}$^*$ \\
\midrule
 M2V+  & DropoutNet & \underline{0.1019} & \underline{0.0443} & \underline{0.0798} & \underline{0.2159} & \underline{0.0782} & \underline{0.1764} & 0.0685 & 0.0126 & 0.0401 \\
       & Heater & 0.0779 & 0.0406 & 0.0687 & 0.1946 & 0.0716 & 0.1658 & \underline{0.0770} & \underline{0.0139} & \underline{0.0481} \\
       & GPatch & \textbf{0.1292(26.79\%)}$^*$ & \textbf{0.0565(27.54\%)}$^*$ & \textbf{0.1172(46.87\%)}$^*$ & \textbf{0.2541(17.69\%)}$^*$ & \textbf{0.0937(19.82\%)}$^*$ & \textbf{0.2204(24.94\%)}$^*$ & \textbf{0.0804(4.42\%)}$^*$ & \textbf{0.0154(10.79\%)}$^*$ & \textbf{0.0535(11.23\%)}$^*$ \\
    \midrule
 LightGCN+ & DropoutNet & 0.0933 & 0.0429 & 0.0795 & 0.2072 & 0.0756 & 0.1796 & -     & -     & - \\
       & Heater & \underline{0.1136} & \underline{0.0534} & \underline{0.1018} & \underline{0.2411} & \underline{0.0873} & \underline{0.2057} & -     & -     & - \\
       & GPatch & \textbf{0.1631(43.57\%)}$^*$ & \textbf{0.0763(42.88\%)}$^*$ & \textbf{0.1411(38.61\%)}$^*$ & \textbf{0.2621(8.71\%)}$^*$ & \textbf{0.0954(9.28\%)}$^*$ & \textbf{0.2296(11.62\%)}$^*$ & -     & -     & - \\
    \bottomrule
 \end{tabular}%
	}}
	\vspace{-1em}
	\label{tab:unified}
	\end{table*}

\begin{table}[ht]
\small
\centering
\caption{Comparison on warm recommendation tasks.}
\vspace{-1em}
\resizebox{1\columnwidth}{!}{
	\setlength{\tabcolsep}{2mm}{
\begin{tabular}{ccccc}
\toprule
   Embeddings   & Models   & CiteULike(NDCG@20) & XING(NDCG@20) & Wechat(NDCG@100) \\
\midrule
MF+   & DropoutNet & 0.0737 & 0.2167 & 0.0394 \\
      & Heater & \underline{0.0934} & \underline{0.2653} & \underline{0.0489} \\
      & GPatch & \textbf{0.1428(52.89\%)}* & \textbf{0.2757(3.92\%)}* & \textbf{0.0533(9.00\%)}* \\
\midrule
M2V+  & DropoutNet & 0.0765 & 0.2211 & 0.0567 \\
      & Heater & \underline{0.0857} & \underline{0.2393} & \underline{0.0629} \\
      & GPatch & \textbf{0.1652(92.77\%)}* & \textbf{0.3157(31.93\%)}* & \textbf{0.0696(10.65\%)}* \\
\midrule
LightGCN+ & DropoutNet & 0.0796 & 0.2506 & - \\
      & Heater & \underline{0.1232} & \underline{0.2930} & - \\
      & GPatch & \textbf{0.1800(46.10\%)}* & \textbf{0.3398(15.97\%)}* & - \\
\bottomrule
\end{tabular}%

 \vspace{-1em}
	}}
	\label{tab:warm}
\end{table}

\begin{table}[ht]
\small
\centering
\vspace{-0.5em}
\caption{Comparison of inferring time.}
\vspace{-1em}
\resizebox{0.7\columnwidth}{!}{
	\setlength{\tabcolsep}{2mm}{

\begin{tabular}{cccc}
\toprule
Model & CiteULike & XING  & WeChat \\
\midrule
DropoutNet & 56 s   & 79 s    & 1259 s\\
Heater & 73 s   & 97 s   & 1483 s\\
GPatch & 43 s   & 64 s   & 1054 s\\
\bottomrule

\end{tabular}%

	}}
	\label{tab:inferring}
\end{table}
	In this section, we aim to answer the following two research questions: RQ1: How does GPatch perform compared with SOTA cold-start recommendation models in terms of effectiveness and efficiency? RQ2: How do the hyper-parameters, such as the dropout ratio $\tau$, affect GPatch's performance?
	
	\paragraph{\textbf{Dataset Description.}}
	We conduct our experiments on two cold-start benchmark datasets~\cite{zhu2020heater}~(CiteULike and XING) and one online commercial dataset ~(WeChat). CiteULike is a user-article dataset, where each article has a 300-dimension tf-idf vector. XING is a user-view-job dataset where each job is described by a 2738-dimension vector. WeChat is a 15-day online user-video dataset, where each video can be represented by their NLP descriptions, a 200-dimension vector. We summarize the three datasets in~\autoref{tab:dataset}. Here we mainly study the cold-start problem of items. For CiteULike and XING, 20\% of items are selected to be the cold-start items. Excluding the cold-start items, 65\% of historical interactions are used for embedding training, while 15\% of historical interactions are used for training GPatch, and the remaining 20\% interactions are used for validation/testing. For WeChat, we split the dataset according to its timeline, where we collect the newly produced videos as the cold-start items. We adopt the popular all-ranking evaluation protocol~\cite{wang2019ngcf,he2020lightgcn,zhu2020heater}. Specifically, we report Precision@20, Recall, and NDCG metrics for the hybrid recommendation task, and we report the NDCG metric for the pure-warm recommendation task.

	\paragraph{\textbf{Baselines \& Embedding Methods.}} We consider two SOTA cold-start recommendation models: DropoutNet~\cite{volkovs2017dropoutnet} and Heater~\cite{zhu2020heater} as our baselines.
	To evaluate the universality of GPatch, we choose three different types of embedding methods:
	\begin{itemize}
		\item MF~\cite{rendle2012bprmf} optimizes the MF~\cite{koren2009MF} with a pairwise ranking loss, which is tailored to learn from implicit feedback.
		\item M2V~\cite{dong2017metapath2vec} formalizes metapath-based random walks on user-item graphs as corpus and then leverages skip-gram~\cite{mikolov2013word2vec} models to compute node embeddings. 
		\item LightGCN~\cite{he2020lightgcn} utilizes GCN layers to aggregate neighbors information, which presents to be currently the most powerful embedding model for user-item recommendation.
	\end{itemize}
	The embedding size is fixed to 200 for all models, and all the embedding methods are implemented with the official codes. We skip LightGCN embeddings on WeChat dataset, since LightGCN reports out of CUDA memory error on a NVIDIA 2080Ti GPU~(11G).
	
    \paragraph{\textbf{Implementation.}}
	In terms of GPatch, we restrict the number of subgraph layers as 3. For computational efficiency, we sample the neighbors to accelerate the training and inference process. We set the sampling size of trajectory size as 25. We use Adam optimizer with a learning rate of 0.001, where the batch size is 1024. The coefficient of $\ell_2$ normalization is set as $1e^{-5}$, and the randomize dropout ratio $\tau$ is set as 0.5. Tanh is used as the activation function. The size of the hidden layers and the output layer is set as 200, while the depth of MLP is set as 2 by default. Moreover, early stopping strategy is performed by observing the AUC scores.
    
    \subsection{Main Results~(Q1)}
	\label{sec:results}
    \autoref{tab:unified} reports the performance comparison of GPatch and baselines on the hybrid recommendation tasks, which evaluates the ability of cold-start models to rank the top-20(100 for WeChat) warm and cold-start items out of all items. We conduct our experiments on three different warm embeddings. From this table, we observe that GPatch statistically outperforms all other baselines on all embeddings and all datasets with $p<0.05$, demonstrating the consistent superiority of our GPatch on hybrid recommendations. This further verifies that separating warm recommendation and cold-start recommendation models can exploit more potentialities from warm/cold-start recommendation models at the same time, resulting in better hybrid recommendation performance.
    
    To verify the pure-warm recommendation performance, in~\autoref{tab:warm}, we present how powerful are GPatch and other baselines in recommending the top-20(100 for WeChat) warm items out of all warm items. From this table, we highlight that GPatch outperforms DropoutNet and Heater with statistically significant margins on all embeddings and all datasets. This further presents the necessity of solving cold-start problems with powerful GCN models.
    
    In terms of efficiency, we present the average inferring time of DropoutNet, Heater, and GPatch on all three datasets. As shown in~\autoref{tab:inferring}, GPatch is more efficient than DropoutNet and Heater. The reason is that GWarmer is as simple as the inner product during inferring, while DropoutNet and Heater still have to generate the embeddings by the MLP function. 
    
    	\begin{figure}[tbp]
		\centering
		\includegraphics[width=\columnwidth, trim=0cm 0cm 0cm 0cm,clip]{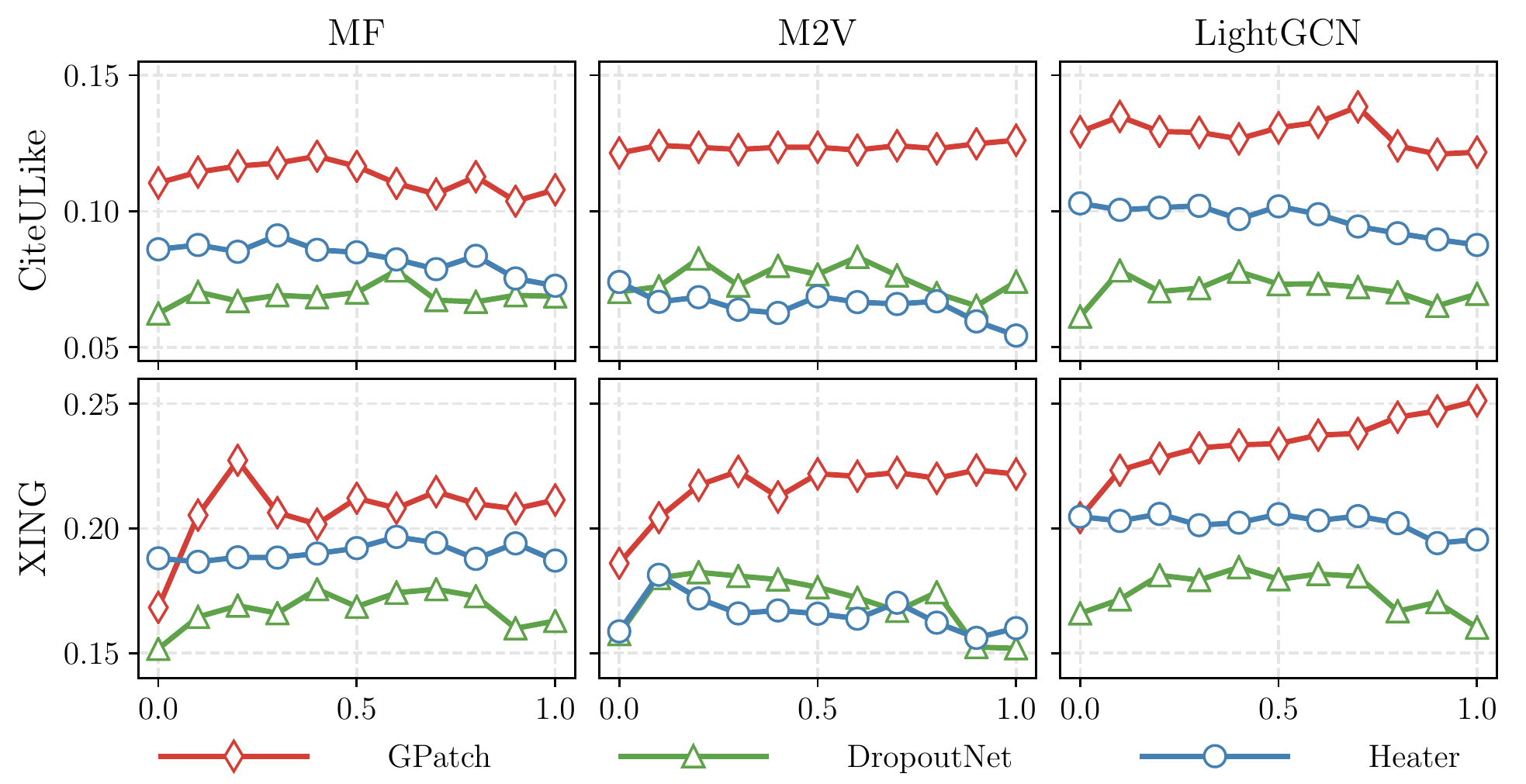}
		\vskip -1em
		\caption{Comparison of different training dropout ratio $\tau$.}
	\vskip -2em	\label{fig:dropout}
	\end{figure}
    \subsection{Ablation Study~(Q2)}
    In~\autoref{fig:dropout}, we compare the hybrid recommendation performance of GPatch, DropoutNet, and Heater by training with different dropout ratio $\tau$. Observing from this table, GPatch consistently outperforms DropoutNet and Heater on most situations. This verifies the consistently superior of GPatch. Especially, since Patching Network is only used for cold-start recommendation. GPatch may achieve the best performance when set $\tau$ as 1.

    \section{Conclusion}
    In this paper, we propose a general cold-start framework named GPatch, which at the first time solves the cold-start problem for GCN models. As a consequence, GPatch is generally powerful on all datasets and embeddings in both the hybrid recommendation task and the warm recommendation task. 
    Besides, the proposed Patching Networks offer researchers an efficient framework to patch powerful warm recommendation models with other patching cold-start models. Since then, future studies can be conducted in 1. designing more powerful patching cold-start models, 2. combing with SOTA warm recommendation models, and 3. investigating specially designed patching models.

	\bibliography{ref}
	\bibliographystyle{unsrt}
	
\end{document}

%% file: math_commands.tex
\usepackage{amsmath,amsfonts,bm}

\def\eqref#1{equation~\ref{#1}}

\def\1{\bm{1}}

\def\vc{{\bm{c}}}

\def\ve{{\bm{e}}}

\def\vx{{\bm{x}}}

\def\mC{{\bm{C}}}

\def\mE{{\bm{E}}}

\def\mT{{\bm{T}}}

\DeclareMathAlphabet{\mathsfit}{\encodingdefault}{\sfdefault}{m}{sl}
\SetMathAlphabet{\mathsfit}{bold}{\encodingdefault}{\sfdefault}{bx}{n}

\def\gD{{\mathcal{D}}}

\def\gG{{\mathcal{G}}}

\def\gI{{\mathcal{I}}}

\def\gL{{\mathcal{L}}}

\def\gP{{\mathcal{P}}}

\def\gR{{\mathcal{R}}}

\def\gT{{\mathcal{T}}}
\def\gU{{\mathcal{U}}}

\def\gW{{\mathcal{W}}}

